\documentclass[floats,floatfix,showpacs,amssymb,prd,twocolumn,superscriptaddress,nofootinbib]{revtex4-1}

\usepackage{amssymb,amsmath,verbatim,mathtools,needspace,enumitem,etoolbox,graphicx,physics,microtype,xspace}

\usepackage[dvipsnames, usenames]{xcolor}

\definecolor{linkcolor}{rgb}{0.0,0.3,0.5}
\usepackage[unicode, colorlinks=true, linkcolor=linkcolor, citecolor=linkcolor, filecolor=linkcolor,urlcolor=linkcolor, pdfusetitle]{hyperref}
\usepackage[all]{hypcap}
\usepackage[T1]{fontenc}
\usepackage[utf8]{inputenc}
\DeclareMathOperator{\sign}{sign}
\usepackage{tikz-cd}
\usepackage{soul}

\def\apj{\rm{ApJ}}
\def\apjl{\rm{ApJ}}

\def\prd{\rm{PRD}}
\def\prl{\rm{PRL}}
\def\prx{\rm{PRX}}

\def\cqg{\rm{CQG}}

\def\ijmpd{\rm{IJMPD}}

\newcommand\ft{FT\xspace}
\newcommand\ift{FT$^{-1}$\xspace}

\newcommand{\caltech}{\affiliation{TAPIR 350-17, California Institute of Technology, 1200 E California Boulevard, Pasadena, CA 91125, USA}}
\newcommand{\montana}{\affiliation{eXtreme Gravity Institute, Department of Physics, Montana State University Bozeman, MT 59717, USA}}
\newcommand{\arizona}{\affiliation{Steward Observatory, University of Arizona, 933 North Cherry Avenue, Tucson, AZ 85721, USA}}

\newcommand{\lisbon}{\affiliation{Centro de Astrof{\'{i}}sica e Gravita{\c{c}}{\~{a}}o -- CENTRA, Departamento de F{\'{i}}sica, Instituto Superior T{\'{e}}cnico -- IST,  Universidade de Lisboa -- UL,  Av.\ Rovisco Pais 1, 1049-001 Lisboa, Portugal}}

\newcommand{\cambridge}{\affiliation{DAMTP, Centre for Mathematical Sciences, University of Cambridge, Wilberforce Road, Cambridge CB3 0WA, UK}}

\begin{document}

\title{Frequency-domain waveform approximants capturing Doppler shifts}

\author{Katie Chamberlain} 
\email{katiechambe@email.arizona.edu}
\montana \caltech \arizona

\author{Christopher J. Moore}
\email{christopher.moore@tecnico.ulisboa.pt}
\lisbon \cambridge

\author{Davide Gerosa} 
\thanks{Einstein Fellow}
\email{dgerosa@caltech.edu}
\caltech

\author{Nicol\'as Yunes} 
\email{nicolas.yunes@montana.edu }
\montana

\date{\today}

\begin{abstract}

Gravitational wave astrophysics has only just begun, and as current detectors are upgraded and new detectors are built, many new, albeit faint, features in the signals will become accessible. 
One such feature %
 is the presence of time-dependent Doppler shifts, generated by the acceleration of the center of mass of the gravitational-wave emitting system.
We here develop a generic method that takes a frequency-domain, gravitational-wave model devoid of Doppler shifts and introduces modifications that incorporate them.
Building upon a perturbative expansion that assumes the Doppler-shift velocity is small relative to the speed of light, the method consists of the inclusion of a single term in the Fourier phase and two terms in the Fourier amplitude.   
We validate the method through matches between waveforms with a Doppler shift in the time domain and waveforms constructed with our method for two toy problems: constant accelerations induced by a distant third body and Gaussian accelerations that resemble a kick profile. 
We find mismatches below $\sim\!10^{-6}$ for all of the astrophysically relevant cases considered, and improve further at smaller velocities.
The work presented here will allow for the use of future detectors to extract new, faint features in the signal from the noise. 

\end{abstract}

\maketitle    

\section{Introduction}

The era of gravitational-wave  (GW) astrophysics has only just begun. The first observations of black hole (BH) mergers~\cite{2016PhRvL.116f1102A,2016PhRvX...6d1015A} and of neutron star mergers~\cite{2017ApJ...848L..12A} have already revealed a 
trove 
of information about both astrophysics~\cite{2018arXiv180605820M,2016ApJ...818L..22A} and extreme gravity~\cite{2016PhRvD..94h4002Y,2016PhRvL.116v1101A}, but they are just the tip of the iceberg. Constructions are already underway to enhance the current network of LIGO-Virgo GW detectors~\cite{2015CQGra..32g4001L,2015CQGra..32b4001A} through the addition of instruments in Japan (KAGRA~\cite{2017arXiv171004823K}) and India (LIGO-India~\cite{2013IJMPD..2241010U}). Future-generation ground based interferometers are currently being planned~\cite{2010CQGra..27s4002P,2017CQGra..34d4001A}, with an expected improvement in sensitivity of more than an order of magnitude.  Moreover, the space-based GW detector LISA is now fully approved and scheduled for launch, opening up the possibility of multi-wavelength GW astrophysics~\cite{2017arXiv170200786A}. These detectors will be much more sensitive than the current Advanced LIGO/Virgo detectors, and will allow us to characterize finer features of loud events and to uncover broad features of quieter signals.     

Among the plethora of finer features that future detectors will be sensitive to, Doppler shifts encoded in the GWs emitted by coalescing compact binaries have the potential to unveil unprecedented (astro)physical information. Doppler shifts naturally arise in a variety of circumstances. For example, if the GW-emitting binary is in the neighborhood of, or in orbit around, a third body~\cite{2011PhRvD..83d4030Y,2017PhRvD..96h3015Y,2017ApJ...834..200M,2017ApJ...834..200M}, its motion in the companion's gravitational potential will be encoded in the emitted GW as a Doppler shift \cite{2018PhRvD..98f4012R,2018arXiv180505335R,2017PhRvD..96f3014I}. Another possibility is for the host galaxy of the binary to possess a peculiar acceleration due to either gravitational attraction to another neighboring galaxy~\cite{2017PhRvD..95d4029B} or the expansion of the Universe~\cite{2001PhRvL..87v1103S,2010PhRvD..81f4018Y,2012PhRvD..85d4047N}. Doppler shifts might also be caused by asymmetric emission of linear momentum in GWs of an isolated binary system~\cite{2016PhRvL.117a1101G}, which imparts a recoil (or ``kick'') velocity to the system's center of mass close to merger~\cite{1961RSPSA.265..109B, 1962PhRv..128.2471P,2007PhRvL..98w1102C, 2007PhRvL..98w1101G,2007PhRvD..76f1502T}. Some proposed modifications of Einstein's General Relativity could also introduce Doppler shifts, for example when fundamental constants of nature become time-dependent~\cite{2010PhRvD..81f4018Y}. 

But not all Doppler shifts are created equal. Galilean 
invariance, a founding block of Einstein's General Relativity, prevents constant velocity Doppler shifts from leaving an observable signature on GWs. Since physics must be the same in all inertial frames, a binary whose center of mass is moving at a constant velocity can always be Lorentz-boosted into a frame that is stationary. This leads to GWs that look functionally identical to the non-boosted ones, but with masses that are rescaled by a constant Doppler shift~\cite{1987GReGr..19.1163K}. Constant Doppler shifts are therefore nearly degenerate with the binary's total mass. This is, for instance, the case of the Universe's cosmological expansion: the phase of the GW emitted by a binary of mass $M$ at redshift $z$ is identical to that of close binary of mass $M(1+z)$. This degeneracy is broken for \emph{time-dependent} Doppler shifts, which do leave an imprint in the emitted signal, as in the examples given above. In an accelerated (i.e. non-inertial) frame, Galilean invariance holds locally, not globally.

In order to detect Doppler-shifts with GW interferometers and extract their (astro)physical origin, one needs waveform models able to capture them. 
Given a generic velocity profile $\mathbf{v}(t)$, can one construct a GW model for a coalescing compact binary that includes the imparted Doppler shift? We show in this paper that this is in fact possible. We show that it can be done entirely in the frequency-domain, and therefore is directly applicable to GW parameter-estimation algorithms. In fact, one can take a standard frequency-domain model in the rest frame of the coalescing binary (non Doppler-shifted) and apply simple analytical modifications to produce an accelerated, Doppler-shifted model. 
Our calculation leverages the stationary phase approximation (SPA) and results in a straightforward ``recipe'' to capture Doppler shifts in any pre-existing frequency-domain GW template. The result we derive consists of the addition of three simple analytic terms in the Fourier amplitude and phase, which only depend on the velocity profile, its first integral (the distance profile), and its first derivative (the acceleration profile). 

We verify the accuracy of our findings for two concrete examples: a constant acceleration profile (meant to represent the effect of a far-away third body) and a generic Gaussian velocity profile centered at merger (meant to mimic the simple recoil model of Ref.~\cite{2016PhRvL.117a1101G}). Our model is then validated through match calculations between waveforms where the Doppler shift is applied explicitly in the time domain (and then discretely Fourier transformed) and waveforms Doppler shifted through our frequency-domain method. 
We find mismatches which improve exponentially as the imparted velocity lies within the perturbative treatment here implemented and are smaller than $\sim\!10^{-6}$ for all the astrophysically relevant examples considered.

The remainder of this paper illustrates the details of the results summarized above. 
Section~\ref{calculations} presents our main calculation and result.
Section~\ref{sec:applications} applies our method to a couple of concrete examples for validation. 
Section~\ref{end} concludes and points to future work. 
Henceforth, 
 we use geometric units in which $c = G = 1$.  

\section{Doppler-shifted gravitational-wave signals}
\label{calculations}

In this section, we first set the stage of the calculation by providing a road-map of the mathematical steps that will be required to the develop our frequency-domain method. We then proceed by introducing our method broken up into two parts: the inverse Fourier transform (\ift) and the forward Fourier transform (\ft) that includes a shift. We conclude this section with a simple recipe that summarizes our frequency-domain method.

\subsection{Setting the stage}
\label{sec:settingup}

\begin{figure*}[t!]
\centering
\includegraphics[clip,width=.75\textwidth]{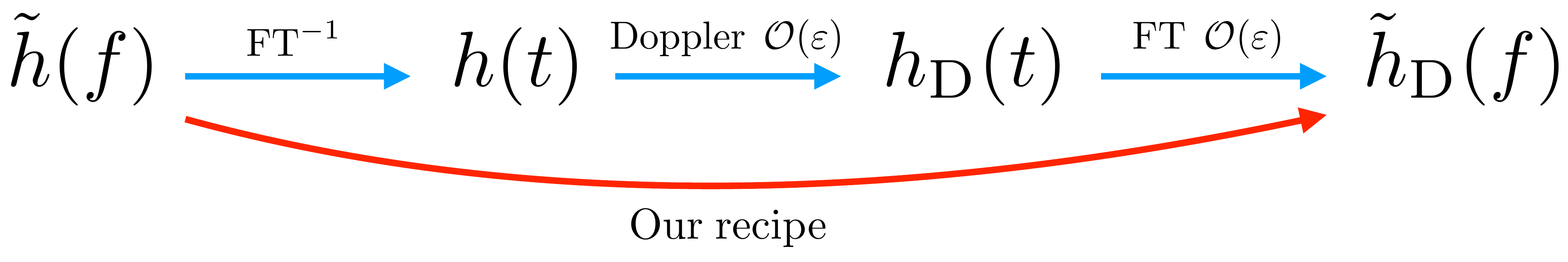}
\caption{Schematic representation of the calculation presented in this paper. The blue arrows indicate the technical steps carried out in Sec.~\ref{calculations}. The red arrow indicates the frequency-domain method we develop to modify a pre-existing frequency-domain waveform approximant to include Doppler shifts.}
\label{hsarrows}
\end{figure*}

The input to our calculation is a pre-existing frequency-domain waveform model. This is characterized by two real functions of frequency, the amplitude  $A(f)$ and the phase $\phi (f)$, which together give the complex frequency-domain strain
\begin{equation}
  \tilde{h}(f) = A(f) \exp\big[i \phi(f)\big] \,. \label{eq:FDwaveform}
\end{equation} 
One should think of this scalar function as the response function of an interferometer due to an impinging GW, i.e.~the contraction of the GW metric perturbation onto the beam patter response tensor.  From this frequency-domain strain, we can compute the time-domain strain $h(t)$ through the \ift of $\tilde{h}(f)$:
\begin{equation}
  h(t) = \int\! A(f)\, \exp\Big\{i\big[\phi( f)-2\pi f t\big]\Big\}\, \textrm{d}f \,. \label{eq:IFT}
\end{equation}

Let us now include the effect of a Doppler shift in the time-domain strain. To do so, we define $v(t)$ to be the velocity imparted to the binary's center of mass projected along the line of sight, i.e. $v(t)=\mathbf{v}(t)\cdot \mathbf{\hat n}$, where $\mathbf{v}$ is the three-velocity of the center of mass and $\mathbf{\hat n}$ is a unit vector directed along the line of sight  pointing from the observer to the source. 
We also define the acceleration (the derivative $a(t) = \mathrm{d}v(t)/\mathrm{d}t$) and the displacement of the source (the integral $d(t) = \int^{t} v(t')\mathrm{d}t'$; the lower limit of integration is degenerate with the distance to the source, it can therefore be set to any convenient reference time).
In the time domain, including a Doppler shift can be phrased as rescaling of the time coordinate.
The relativistic Doppler shift could be used here; however, in the next section it will be necessary to expand to leading order in the velocity, here we preempt this by using the simpler non-relativistic formula and rescale
$t\rightarrow t_{\textrm{D}}(t)$, where
\begin{align}
 \textrm{d}t_{\textrm{D}} &= \textrm{d}t \times \big[1+\varepsilon \, v(t)\big] \label{eq:redshift} \\
t_{\textrm{D}}(t) &= t + \varepsilon \, d(t) \,.
 \label{eq:redshifttime}
\end{align}
This classical Doppler shift formula requires $v(t) \ll 1$ is a small perturbation parameter, and $\varepsilon$ is a book-keeping parameter introduced to label the order of the perturbation. This defines the new Doppler shifted time-domain waveform
\begin{equation}
  h_{\textrm{D}}(t) = h\big[t_{\textrm{D}}(t)\big] \,.
\end{equation}

The strategy of our calculation is shown schematically in Fig.~\ref{hsarrows} and consists of the following steps: 
\begin{enumerate}
\item[(1)] Perform a \ift of the input frequency-domain waveform; 
\item[(2)] Implement the Doppler shift in the time domain on the time-domain waveform that resulted from (1);
\item[(3)] Compute a \ft at leading order in $\varepsilon$ to transform the Doppler-shifted, time-domain waveform back to the frequency domain. 
\end{enumerate} 
We perform both the \ift and the \ft integrals analytically using the SPA. This allows us to obtain a simple analytical prescription to transform directly from $\tilde{h}(f)$ to $\tilde{h}_{\textrm{D}}(f)$ for any given velocity profile $v(t)$. Readers interested in the final recipe can skip the next three subsections and proceed to  Sec.~\ref{recipe} where our main finding is presented concisely.

 \subsection{Inverse Fourier transform}
\label{iftsec}
 
First, we tackle the \ift to transform $\tilde{h}(f)$ into ${h}(t)$, i.e. the integral in Eq.~(\ref{eq:IFT}). We will assume that the amplitude of the integrand varies much more slowly than the phase, so that the SPA is valid. Due to the highly oscillatory nature of the integrand the result is dominated by contributions close to some critical frequency $f_{\textrm{s}}(t)$ where the phase has a stationary point. To simplify the notation, from now on we drop the argument of $f_{\textrm{s}}(t)$ and only indicate it explicitly in some key equations. It is important to remember though that $f_{\rm s}$ is a function of time, and not a fixed frequency value.

Let us now use the above to simplify the expressions that appear in the integrand of Eq.~\eqref{eq:IFT}. We begin by Taylor expanding the phase of the integrand 
\begin{equation}
X(f,t)\equiv\phi(f)-2\pi f t
\end{equation}
to quadratic order about the critical frequency to find
\begin{align}
  X(f,t) \approx &\, \big[ \phi(f_{\textrm{s}})-2\pi f_{\textrm{s}}t\big]  %
   + \big(f-f_{\textrm{s}}\big)\bigg[ \frac{\textrm{d}\phi}{\textrm{d}f}\Big|_{f_{\textrm{s}}} -2\pi t\bigg]\nonumber \\ 
  &  + \frac{1}{2}\big(f-f_{\textrm{s}}\big)^{2}\, \frac{\textrm{d}^2\phi}{\textrm{d}f^2}\Big|_{f_{\textrm{s}}} \,.%
\label{eq:TaylorExp} 
\end{align}
Similarly, Taylor expanding the amplitude to zeroth order we find
\begin{align}
  A(f) \approx &\,A(f_{\textrm{s}})\,.
\end{align}
The critical frequency $f_{\textrm{s}}(t)$ is the function of time at which of the phase of the integrand $X(f,t)$ is slowly-varying, i.e. the function that sets the first-order term in the Taylor expansion of Eq.~(\ref{eq:TaylorExp}) to zero, namely 
\begin{equation} \label{eq:SPAcondition}
\frac{\textrm{d}\phi}{\textrm{d}f}\Big|_{f_{\textrm{s}}}=2\pi t \,,
\end{equation}
which is sometimes called the SPA condition \cite{Maggiore:1900zz,1978amms.book.....B}.

With this at hand, the \ift of Eq.~(\ref{eq:IFT}) becomes
\begin{align}
  h(t) & \approx A\big(f_{\textrm{s}}\big)\exp\bigg\{i\Big[\phi\big(f_{\textrm{s}}\big)-2\pi f_{\textrm{s}}t\Big]\bigg\} \nonumber \\
  & \times\int\!\,\exp\bigg[\frac{i}{2}\big(f-f_{\textrm{s}}\big)^{2}\ \frac{\textrm{d}^2\phi}{\textrm{d}f^2}\Big|_{f_{\textrm{s}}} \bigg]\, \textrm{d}f \,,  \label{eq:TEMP} \\
  h(t) & \approx A\big(f_{\textrm{s}}\big)\exp\bigg\{i\Big[\phi\big(f_{\textrm{s}}\big)-2\pi f_{\textrm{s}}t\Big]\bigg\} \nonumber \\
  & \times\! \sqrt{\frac{\pi}{\big|{\rm d}^2\phi / {\rm d} f^2 |_{f_{\textrm{s}}} \big|}} \bigg[1 + i\,\sign\bigg( \frac{\textrm{d}^2\phi}{\textrm{d}f^2}\Big|_{f_{\textrm{s}}} \bigg)\bigg] \,, \label{eq:TDSPAexpression}
\end{align}
where in the second line we have evaluated the standard Gaussian integral. Now that the time domain waveform has been obtained, the next step is to Doppler shift it, and finally to perform a \ft to return to the frequency domain. It is illustrative, however, to first skip the Doppler-shift  step and just to apply the \ft. This warm-up exercise will turn out to be extremely useful to understand the more complex result presented in Sec.~\ref{sec:kickedcase}.

\subsection{Forward Fourier transform: no shift}
\label{sec:unkickedcase}

The frequency-domain waveform is calculated by performing a \ft on the time-domain waveform, i.e.  
\begin{equation}
  \tilde{h}(f) = \int\!\, h(t) \exp(2\pi if t)\, \textrm{d}t \,, \label{eq:FFT}
\end{equation} 
which, upon using Eq.~(\ref{eq:TDSPAexpression}), becomes
\begin{align}
  \tilde{h}(f) & = \! \int\!\, Z(t) \exp \left[i \; Y(t)\right] \textrm{d}t\,,
  \label{eq:FFTdef1}
\end{align}
where we have defined the phase and amplitude
\begin{align}
  Z(t) & = A\big(f_{\textrm{s}}\big) \sqrt{\frac{\pi}{\big|{\rm d}^2\phi / {\rm d} f^2 |_{f_{\textrm{s}}} \big|}} \bigg[1 + i\,\sign\bigg( \frac{\textrm{d}^2\phi}{\textrm{d}f^2}\Big|_{f_{\textrm{s}}} \bigg)\bigg]\,,
  \\
  Y(t) & =  \phi(f_{\textrm{s}})+ 2\pi (f- f_{\textrm{s}}) t \,. 
\end{align}
Let us now solve this integral using the SPA again. As before, the integral is dominated by contributions close to some critical time $t_{\textrm{s}}(f)$, which is a function of frequency; once again, we will drop the argument, i.e. the explicit frequency dependence, from now on to simplify notation.

Let us now carry out our Taylor expansions to simplify the integrand above. Taylor expanding the phase $Y(t)$ 
to quadratic order about $t_{\textrm{s}}$, we find
\begin{align}
  &Y(f,t) \approx 
   \Bigg\{ \phi\Big[f_{\textrm{s}}\big(t_{\textrm{s}}\big)\Big] + 2\pi \Big[ f - f_{\textrm{s}}\big(t_{\textrm{s}}\big) \Big]  t_{\textrm{s}} \Bigg\} 
 \nonumber \\
  & \;+ \big(t-t_{\textrm{s}}\big) \Bigg\{ \frac{\textrm{d}f_{\textrm{s}}}{\textrm{d}t}\Big|_{t_{\textrm{s}}}
   \bigg[  \frac{\textrm{d}\phi}{\textrm{d}f}\Big|_{f_{\textrm{s}}(t_{\rm s})} -2\pi t_{\textrm{s}}\bigg]
   + 2\pi \Big[f -f_{\textrm{s}}\big(t_{\textrm{s}}\big)\Big] 
    \Bigg\}  \nonumber \\
  &  \; + \frac{1}{2}\big(t-t_{\textrm{s}}\big)^{2} \Bigg\{ 
  \frac{\textrm{d}^{2}f_{\textrm{s}}}{\textrm{d}t^{2}}\Big|_{t_{\textrm{s}}} \bigg[  \frac{\textrm{d}\phi}{\textrm{d}f}\Big|_{f_{\textrm{s}}(t_{\rm s})} -2\pi t_{\textrm{s}}\bigg] 
  \notag \\ 
  &  \qquad \qquad \;\;\;\;
  - 4\pi\frac{\textrm{d}f_{\textrm{s}}}{\textrm{d}t}\Big|_{t_{\textrm{s}}} + \frac{\textrm{d}f_{\textrm{s}}}{\textrm{d}t}\Big|^{\;2}_{t_{\textrm{s}}} \;
  \frac{\textrm{d}^2\phi}{\textrm{d}f^2}\Big|_{f_{\textrm{s}}(t_s)} 
  \Bigg\} \,.
\end{align}
In this case, the SPA condition requires that the second term in this expansion vanishes.
Simplifying this through the \ift SPA condition of Eq.~(\ref{eq:SPAcondition}), the new SPA condition is equivalent to 
\begin{align}\label{eq:newSPAconditionUN}
  f_{\textrm{s}}\big[t_{\textrm{s}}(f)\big]  =  f\,.
\end{align}
Let us stress once more that in our notation $t$ and $f$ are scalar quantities (time and frequency respectively), while quantities with subscript, like $t_{\textrm{s}}$ and $f_{\textrm{s}}$, are functions.
The two SPA conditions in Eqs.~(\ref{eq:SPAcondition}) and (\ref{eq:newSPAconditionUN}) jointly imply that 
\begin{align} \label{eq:t0off}
  t_{\textrm{s}}(f) = \frac{1}{2\pi}\frac{\textrm{d}\phi}{\textrm{d}f}\,,
\end{align}
which is the familiar \ft SPA expression of a stationary point for the critical time $t_{\rm s}$.
Using Eq.~(\ref{eq:newSPAconditionUN}), the expansion for the phase simplifies to
\begin{align}
  Y(f,t) \, \approx \, \phi(f) - \pi\big(t-t_{\textrm{s}}\big)^{2} \,\frac{\textrm{d}f_{\textrm{s}}}{\textrm{d}t}\Big|_{t_{\textrm{s}}} %
\end{align}
where we used the relationship 
\begin{equation}
\frac{\textrm{d}^2\phi}{\textrm{d}f^2}\Big|_{f_{\textrm{s}}}\,  \frac{\textrm{d}  f_{\textrm{s}}}{\textrm{d}t}=2\pi
\label{secondderphipp}
\end{equation}
which can be derived from Eq.~(\ref{eq:SPAcondition}). We note for later that  
evaluating Eq.~(\ref{secondderphipp}) at $t=t_{\textrm{s}}(f)$ implies that 
\begin{equation}
 \sign\left[ \frac{\textrm{d}^2\phi}{\textrm{d}f^2}\Big|_{f_{\textrm{s}}(t_{\rm s})}\right] 
= 
\sign \left[ \frac{\textrm{d}  f_{\textrm{s}}}{\textrm{d}t}\Big|_{t_{\rm s}}  \right].
 \label{signeq}
 \end{equation}
 
The amplitude of the integrand in Eq.~(\ref{eq:FFTdef1}) can be easily expanded to zeroth order about $t_{s}$ to find
\begin{align}
Z(f,t)&\approx A\big[f_{\textrm{s}}(t_{\rm s})\big] \sqrt{\frac{\pi}{\big|{\rm d}^2\phi / {\rm d} f^2 |_{f_{\textrm{s}(t_{\rm s})}} \big|}}
\nonumber \\& \times
\bigg[1 + i\,\sign\bigg( \frac{\textrm{d}^2\phi}{\textrm{d}f^2}\Big|_{f_{\textrm{s}}(t_{\rm s})} \bigg)\bigg]\,.
\end{align} 
 The result can then be simplified using Eq.~(\ref{eq:newSPAconditionUN}) to read
\begin{align}
Z(f,t) \approx &\, A(f) \sqrt{\frac{\pi}{\big|{\rm d}^2\phi / {\rm d} f^2  \big|}}
\bigg[1 + i\,\sign\bigg( \frac{\textrm{d}^2\phi}{\textrm{d}f^2}\bigg)\bigg] \,.
\label{eq:Z-no-v}
\end{align}

With the expansions for the amplitude and phase in hand, the \ft of Eq.~(\ref{eq:FFTdef1}) becomes
\begin{align}
  \tilde{h}(f) 
 & = 
 A(f) \sqrt{\frac{\pi}{\left| {\textrm{d}^2\phi}/{\textrm{d}f^2} \right|}} \bigg[1+i\,\sign\bigg( \frac{\textrm{d}^2\phi}{\textrm{d}f^2} \bigg)\bigg] \exp\big[i\phi(f)\big] \nonumber \\
  & \times  \sqrt{\frac{1}{2 \left|
   {\textrm{d}f_{\textrm{s}}}/{\textrm{d}t}\big|_{t_{\textrm{s}}} \right|}} 
  \left[1 - i\,\sign\left(\frac{\textrm{d}f_{\textrm{s}}}{\textrm{d}t}\Big|_{t_{\textrm{s}}} \right) \right] \,.
  \label{tildehfnoshift}
\end{align}
Using Eqs.~(\ref{eq:newSPAconditionUN}), (\ref{secondderphipp}) and (\ref{signeq}),
this simplifies to
\begin{equation}
  \tilde{h}(f) = A(f)\exp\big[i \phi(f)\big] \,. 
\end{equation}
which is equal to our starting point in Eq.~(\ref{eq:FDwaveform}).

What have we shown here? We first performed an \ift using the SPA. We then took a \ft also using the SPA, which resulted in recovering the frequency-domain waveform we started with. We therefore confirmed that the approximations made in the SPA do not undermine the Fourier inversion theorem, as expected. This is an important point, because it implies that our recipe for Doppler shifting a waveform leaves the input approximant untouched in the 
$v(t)=0$ case.

\subsection{Forward Fourier transform: Doppler shift}\label{sec:kickedcase}

Let us now repeat the same \ft calculation as in Sec.~\ref{sec:unkickedcase} but now including a Doppler shift in the time-domain waveform. 
We return to the time-domain waveform of Eq.~(\ref{eq:TDSPAexpression}), and as described above in Sec.~\ref{sec:settingup}, we now need to first make the the substitution ${t\rightarrow t_{\textrm{D}}(t)}$ and then perform a \ft. Put another way, the Doppler-shifted time-domain waveform is 
\begin{align}
&h_{\textrm{D}}(t) \approx A\big[f_{\textrm{s}}(t_{\textrm{D}})\big]  \exp\bigg(i\Big\{\phi\big[f_{\textrm{s}}(t_{\textrm{D}})\big] 
-  2\pi f_{\textrm{s}}(t_{\textrm{D}})t_{\textrm{D}}\Big\}\!\bigg)
\notag \\
&\times\! \sqrt{\frac{\pi}{\Big|
{\textrm{d}^2\phi}/{\textrm{d}f^2}\big|_{f_{\textrm{s}} (t_{\rm D})}
\Big|}} 
\;\bigg[1\!+\!i\,\sign\bigg(\frac{\textrm{d}^2\phi}{\textrm{d}f^2}\Big|_{f_{\textrm{s}} (t_{\rm D}) }\bigg)\bigg] , \label{eq:kickedTDwave}
\end{align}
where $t_{\textrm{D}}\equiv t_{\rm D}(t) =t+\varepsilon \; d(t)$ as defined in Eq.~(\ref{eq:redshifttime}), and we now need to compute the \ft of Eq.~(\ref{eq:kickedTDwave}), namely
\begin{align}
 \tilde{h}_{\textrm{D}}(f) \! &= \!\! \int\!\, A\big[f_{\textrm{s}}(t_{\textrm{D}})\big] 
 \notag \\
&\times \sqrt{\frac{\pi}{\Big|
{\textrm{d}^2\phi}/{\textrm{d}f^2}\big|_{f_{\textrm{s}} (t_{\rm D})}
\Big|}} 
\;\bigg[1\!+\!i\,\sign\bigg(\frac{\textrm{d}^2\phi}{\textrm{d}f^2}\Big|_{f_{\textrm{s}} (t_{\rm D}) }\bigg)\bigg]\notag\\
 &
 \times  \exp\bigg(i\Big\{\phi\big[f_{\textrm{s}}(t_{\textrm{D}})\big] 
+  2\pi \big[ ft - f_{\textrm{s}}(t_{\textrm{D}})t_{\textrm{D}}\big]\Big\}\!\bigg) \textrm{d}t
\notag \,. \\
\label{eq:kickedTDwave}
\end{align}

The calculation closely mirrors the warm up exercise of Sec.~\ref{sec:unkickedcase}. First, the phase of the integrand, $\mathcal{Y}(f,t)$, is expanded to quadratic order about the critical time $t_{\textrm{s}}(f)$:
\begin{align}
\mathcal{Y}(f,t)&\equiv\phi\big[f_{\textrm{s}}(t_{\textrm{D}})\big] 
+  2\pi \big[ ft - f_{\textrm{s}}(t_{\textrm{D}})t_{\textrm{D}}\big] \\
  \mathcal{Y}(f,t)  &\approx  \mathcal{Y}^{(0)}(f,t)  + \big(t-t_{\textrm{s}}\big) \; \mathcal{Y}^{(1)}(f,t)  
  \nonumber \\
  &+ \frac{1}{2}\big(t-t_{\textrm{s}}\big)^{2} \;\mathcal{Y}^{(2)}(f,t) \,,  
\end{align}
where the following coefficients in the Taylor expansion have been defined,
\allowdisplaybreaks[4]
\begin{widetext}
\begin{align}
  \mathcal{Y}^{(0)}(f,t) &\equiv \phi\Big\{f_{\textrm{s}}\big[t_{\textrm{D}}(t_{\textrm{s}})\big]\Big\} + 2\pi \Big\{ f t_{\textrm{s}}  - f_{\textrm{s}}\big[t_{\textrm{D}}(t_{\textrm{s}})\big]t_{\textrm{D}}(t_{\textrm{s}})\Big\} 
  \nonumber \\  
   \mathcal{Y}^{(1)}(f,t) &\equiv \frac{\textrm{d}t_{\textrm{D}}}{\textrm{d}t}\Big|_{t_{\textrm{s}}} \frac{\textrm{d}f_{\textrm{s}}}{\textrm{d}t}\Big|_{t_{\textrm{D}}\left(t_{\textrm{s}}\right)}\frac{\textrm{d}\phi}{\textrm{d}f}\Big|_{f_{\textrm{s}}[t_{\textrm{D}}(t_{\textrm{s}})]} + 2\pi \bigg\{ f - f_{\textrm{s}}\big[t_{\textrm{D}}(t_{\textrm{s}})\big] \frac{\textrm{d}t_{\textrm{D}}}{\textrm{d}t}\Big|_{t_{\textrm{s}}}  \bigg\}
 - 2\pi \frac{\textrm{d}t_{\textrm{D}}}{\textrm{d}t}\Big|_{t_{\textrm{s}}}  \frac{\textrm{d}f_{\textrm{s}}}{\textrm{d}t}\Big|_{t_{\textrm{D}}\left(t_{\textrm{s}}\right)} t_{\textrm{D}}(t_{\textrm{s}})    
  \nonumber \\  
  \mathcal{Y}^{(2)}(f,t) &\equiv  
  \Bigg\{ \frac{\textrm{d}^{2}f_{\textrm{s}}}{\textrm{d}t^{2}}\Big|_{t_{\textrm{D}}\left(t_{\textrm{s}}\right)} \bigg[\frac{\textrm{d}\phi}{\textrm{d}f}\Big|_{f_{\textrm{s}}[t_{\textrm{D}}(t_{\textrm{s}})]} -2\pi t_{\textrm{D}}(t_{\textrm{s}}) \bigg] - 4\pi\frac{\textrm{d}f_{\textrm{s}}}{\textrm{d}t}\Big|_{t_{\textrm{D}}\left(t_{\textrm{s}}\right)}  
  + \frac{\textrm{d}f_{\textrm{s}}}{\textrm{d}t}\Big|^{\;2}_{t_{\textrm{D}}\left(t_{\textrm{s}}\right)} \frac{\textrm{d}^2\phi}{\textrm{d}f^2}\Big|_{f_{\textrm{s}}[t_{\textrm{D}}(t_{\textrm{s}})]}
\Bigg\}
\frac{\textrm{d}t_{\textrm{D}}}{\textrm{d}t}\Big|^{\;2}_{t_{\textrm{s}}}
  \nonumber \\ & \quad\;  + 
  \bigg\{ \frac{\textrm{d}f_{\textrm{s}}}{\textrm{d}t}\Big|_{t_{\textrm{D}}\left(t_{\textrm{s}}\right)}\frac{\textrm{d}\phi}{\textrm{d}f}\Big|_{f_{\textrm{s}}[t_{\textrm{D}}(t_{\textrm{s}})]} - 2\pi f_{\textrm{s}}\big[t_{\textrm{D}}(t_{\textrm{s}})\big] -2\pi \frac{\textrm{d}f_{\textrm{s}}}{\textrm{d}t}\Big|_{t_{\textrm{D}}\left(t_{\textrm{s}}\right)} t_{\textrm{D}}(t_{\textrm{s}})\bigg\}  \frac{\textrm{d}^{2}t_{\textrm{D}}}{\textrm{d}t^{2}}\Big|_{t_{\textrm{s}}} \,.
  \label{eq:GIANTmess}
\end{align}
\end{widetext}
As above, the SPA condition is equivalent to imposing 
\begin{equation}
\mathcal{Y}^{(1)}(f,t)=0
\end{equation}
and using Eq.~(\ref{eq:SPAcondition}), one obtains
\begin{align}\label{eq:newSPAconditionKick}
  f_{\textrm{s}}\Big\{t_{\textrm{D}}\big[t_{\textrm{s}}(f)\big]\Big\} \; \frac{\textrm{d}t_{\textrm{D}}}{\textrm{d}t}\Big|_{t_{\textrm{s}}(f)} = f \,.
\end{align}
Note that this correctly reduces to Eq.~(\ref{eq:newSPAconditionUN}) if the function $t_D(t) = 1$; i.e.\ if $v(t)=0$ at all times.

With the \ft SPA condition in Eq.~(\ref{eq:newSPAconditionKick}) at hand, we can now simplify the remaining terms in the Taylor expansion of the phase. Let us first use Eq.~(\ref{eq:newSPAconditionKick}) to eliminate $f_{\textrm{s}}[t_{\textrm{D}}(t_{\textrm{s}})]$, so that the constant term $\mathcal{Y}^{(0)}(f,t)$ becomes
\begin{align}
  \mathcal{Y}^{(0)}(f,t) &= \phi\left(\frac{f}{ {\textrm{d}t_{\textrm{D}}}/{\textrm{d}t}\big|_{t_{\textrm{s}}} }\right) 
  + 2\pi f \left[  t_{\textrm{s}} -  \frac{t_{\textrm{D}}\big(t_{\textrm{s}}\big)}{ {\textrm{d}t_{\textrm{D}}}/{\textrm{d}t}\big|_{t_{\textrm{s}}} } \right]  \,.
\end{align}
Substituting for $t_{\textrm{D}}(t_{\textrm{s}})$ and $\textrm{d}t_{\textrm{D}}/\textrm{d}t|_{t_{\textrm{s}}}$ from Eqs.~(\ref{eq:redshift}) and (\ref{eq:redshifttime}) respectively, and performing a Taylor expansion in powers of $\varepsilon$ to first order gives
\begin{align}\label{eq:NEWEQREF1}
    \mathcal{Y}^{(0)}(f,t)  &= \,\phi(f) - f \varepsilon v(t_{\textrm{s}})\frac{{\rm d}\phi}{{\rm d}f} 
    \nonumber \\ & 
   + 2\pi f \varepsilon \big[ t_{\textrm{s}}  v(t_{\textrm{s}} -d(t_{\textrm{s}})\big] + {\cal{O}}(\varepsilon^{2})\,.
\end{align}
As stressed above, the leading-order \ft SPA condition of  Eq.~(\ref{eq:newSPAconditionKick}) is given by Eq.~(\ref{eq:newSPAconditionUN}), i.e. $f_{\textrm{s}}[t_{\textrm{s}}(f)] = f + \mathcal{O}(\varepsilon)$. Combining this with the \ift SPA condition in Eq.~(\ref{eq:SPAcondition}) yields ${\rm d}\phi/{\rm d}f =2\pi t_{\textrm{s}}(f)+\mathcal{O}(\varepsilon)$. We can therefore further simplify Eq.~(\ref{eq:NEWEQREF1})  to
\begin{align}
\mathcal{Y}^{(0)}(f,t)  = \phi(f) - 2\pi \; f \; \varepsilon \; d\big(t_{\textrm{s}}) + {\cal{O}}(\varepsilon^{2})\,.
\end{align}

The quadratic term in Eq.~(\ref{eq:GIANTmess}) $\mathcal{Y}^{(2)}(f,t)$  may be simplified using  Eq.~(\ref{eq:SPAcondition}) and its derivative with respect to $t$ evaluated at $t=t_{\textrm{D}}(t_{\textrm{s}})$:
\begin{align}
\mathcal{Y}^{(2)}(f,t) =  
  -2\pi\bigg\{ \frac{\textrm{d}f_{\textrm{s}}}{\textrm{d}t}\Big|_{t_{\textrm{D}}\left(t_{\textrm{s}}\right)} \frac{\textrm{d}t_{\textrm{D}}}{\textrm{d}t}\Big|^{\,2}_{t_{\textrm{s}}} 
+ f_{\textrm{s}}\big[t_{\textrm{D}}\big(t_{\textrm{s}}\big)\big] \frac{\textrm{d}^{2}t_{\textrm{D}}}{\textrm{d}t^{2}}\Big|_{t_{\textrm{s}}} \bigg\}.
\end{align}
Using the \ft SPA condition in Eq.~(\ref{eq:newSPAconditionKick}) to eliminate $f_{\textrm{s}}[t_{\textrm{D}}(t_{\textrm{s}})]$, one then finds
\begin{align}
 \mathcal{Y}^{(2)}(f,t) \! =\! & 
  -2\pi \!\left[ \frac{\textrm{d}f_{\textrm{s}}}{\textrm{d}t}\Big|_{t_{\textrm{D}}\left(t_{\textrm{s}}\right)} \frac{\textrm{d}t_{\textrm{D}}}{\textrm{d}t}\Big|^{\,2}_{t_{\textrm{s}}} 
  + \left(\frac{ {\textrm{d}^{2}t_{\textrm{D}}}/{\textrm{d}t^{2}}\Big|_{t_{\textrm{s}}} }{ {\textrm{d}t_{\textrm{D}}}/{\textrm{d}t}\Big|_{t_{\textrm{s}}} }\right)f \right].
\end{align}
Substituting $\textrm{d}t_{\textrm{D}}/\textrm{d}t = 1+\varepsilon \; v(t)$ and $\textrm{d}^{2}t_{\textrm{D}}/\textrm{d}t^{2} = \varepsilon \; a (t)$, and performing a Taylor expansion to first order in $\varepsilon$  gives
\begin{align}
\mathcal{Y}^{(2)}(f,t)  &= -2\pi \Big\{ \big[1+2\varepsilon v(t_{\rm s})\big]     \frac{\textrm{d}f_{\textrm{s}}}{\textrm{d}t}\Big|_{t_{\textrm{D}}\left(t_{\textrm{s}}\right)}  
\nonumber \\ 
&+
 f \varepsilon a ( t_{\textrm{s}})   + {\cal{O}}(\varepsilon^{2}) \Big\}\,. \label{eq:third_finaldef}
\end{align}

Finally, let us also expand the amplitude $\mathcal{Z}(f,t)$ from Eq.~(\ref{eq:kickedTDwave}) to zeroth order in time
\begin{align}
\mathcal{Z}(f,t) \approx &\,  A\Big(f_{\textrm{s}}\big[t_{\textrm{D}}(t_{\textrm{s}})\big]\Big) \sqrt{\frac{\pi}{\left|{\rm d}^2\phi/{\rm d}f^2\big|_{f_{\textrm{s}}[t_{\textrm{D}}(t_{\textrm{s}})]}\right|}} \nonumber \\ &\, \times \bigg[1+ i\,\sign\bigg( \frac{{\rm d}^2\phi}{{\rm d}f^2}\Big|_{f_{\textrm{s}}[t_{\textrm{D}}(t_{\textrm{s}})]} \bigg)\bigg] \,.
\label{zetaamplitude}
\end{align}
Notice, as before, that this results reduces to that of Eq.~\eqref{eq:Z-no-v} when $v(t) = 0$. 

We now have all the ingredients to perform the \ft integral of  Eq.~(\ref{eq:FFT}) in the SPA within a small $\varepsilon$ expansion:
\begin{align}
  \tilde{h}_{\textrm{D}}(f) 
  &= \, \mathcal{Z}(f,t)   \frac{1+i\,\sign\big[\mathcal{Y}^{(2)}(f,t)\big]}{\sqrt{\big|\mathcal{Y}^{(2)}(f,t)\big|/\pi}} 
  \notag \\ &\times
  \exp\Big\{
  i\big[ \phi(f)-2\pi f \varepsilon d(t_{\textrm{s}}) \big]
  \Big\}
  \,,
\end{align}
where $\mathcal{Z}(f,t)$ and $\mathcal{Y}^{(2)}(f,t)$ are given by Eqs.~(\ref{eq:third_finaldef}) and  (\ref{zetaamplitude}) and we carried out a standard Gaussian integral and used Eq.~(\ref{signeq}). The expression above may be further simplified using the SPA conditions and by Taylor expanding in powers of $\varepsilon$ to find
\begin{align} \label{eq:notsoFINAL}
  \tilde{h}_{\textrm{D}}(f)  &= \, \bigg\{
   A(f) - \left[A(f)+f\frac{{\rm d}A}{{\rm d}f}\right] \varepsilon v(t_{\textrm{s}}) \nonumber \\ &- \frac{f A(f)}{4\pi} \frac{{\rm d}^2 \phi}{{\rm d}f^2} \varepsilon a(t_{\textrm{s}}) + {\cal{O}}(\varepsilon^{2})
      \bigg\} 
      \nonumber \\ &
      \times\exp\bigg\{
  i\Big[ \phi(f)-2\pi f \varepsilon d(t_{\textrm{s}}) + {\cal{O}}(\varepsilon^{2}) \Big] 
  \bigg\} \,.
\end{align}
Finally, at $\mathcal{O}(\varepsilon)$ we can replace $t_{\textrm{s}}$ with  $({\rm d}\phi/{\rm d}f)/2\pi$ and obtain our final result:
\begin{align} \label{eq:FINAL}
  \tilde{h}_{\textrm{D}}(f)  &= \, \bigg\{
   A(f) - \left[A(f)+f\frac{{\rm d}A}{{\rm d}f}\right] \varepsilon v\left( \frac{{\rm d}\phi/{\rm d}f}{2\pi} \right) \nonumber \\ &- \frac{f A(f)}{4\pi} \frac{{\rm d}^2 \phi }{{\rm d}f^2} \varepsilon a\left( \frac{{\rm d}\phi/{\rm d}f}{2\pi} \right) + {\cal{O}}(\varepsilon^{2})
      \bigg\} 
      \nonumber \\ &
      \times\exp\bigg\{
  i\Big[ \phi(f)-2\pi f \varepsilon d\left( \frac{{\rm d}\phi/{\rm d}f}{2\pi} \right) + {\cal{O}}(\varepsilon^{2})\Big] 
  \bigg\} \,.
\end{align}
The book keeping parameter $\varepsilon$  is no longer needed and will be set to unity henceforth. Equation~(\ref{eq:FINAL}) is an explicit expression for the Doppler-shifted waveform in the frequency domain.

As a sanity check, we can examine this result in the simple case of constant velocity which, as is well known, is degenerate with the total mass of the source: $d(t)=vt$, $v(t)=v$, $a(t)=0$.  A constant velocity gives the simple Doppler-shift $h_{\rm D}(t)=h[t(1+v)]$. In the frequency domain this becomes
\begin{align} 
  \tilde{h}_{\rm D}(f)&=\int\mathrm{d}t\,h[t(1+v)]\exp(2\pi i ft) \nonumber \\
  &=\big[1-v+\mathcal{O}(v^{2})\big]\tilde{h}\Big\{f\big[1-v+\mathcal{O}(v^{2})\big]\Big\} \,. \label{eq:SanityCheck}
\end{align}
In the second line we have changed integration variables to $t'=t(1+v)$ and used the definition of $\tilde{h}(f)$ in Eq.~(\ref{eq:FFT}).
From Eq.~(\ref{eq:FINAL}) one can see that 
\begin{align}
  \tilde{h}_{\textrm{D}}(f)  = &\bigg[ (1-v)A(f) + fv\frac{{\rm d}A}{{\rm d}f} + \mathcal{O}(v^{2}) \bigg] \nonumber \\
  &\times \exp\bigg\{ i \bigg[ \phi(f) - fv\frac{{\rm d}\phi}{{\rm d}f} + \mathcal{O}(v^{2}) \bigg] \bigg\} \,.
  \label{eq:SanityCheck2}
\end{align}
Equations~(\ref{eq:SanityCheck}) and (\ref{eq:SanityCheck2}) and can be put into agreement at the required order in $v$ using the expansion
\begin{equation}
X[f(1-v)]=X(f)-fv \frac{{\rm d}X}{{\rm d}f}+\mathcal{O}(v^{2})\,.
\end{equation}
\subsection{A simple recipe}
\label{recipe}

To summarize, here is a simple recipe to add a Doppler-shift  to a GW waveform entirely in the frequency domain (Fourier transforms are not required!):

\begin{enumerate}
\item Start with an unkicked waveform model in the frequency domain:
\begin{equation}
\tilde{h}(f) = A(f)\exp\big[i \phi(f)\big]\,,
\end{equation}
and a velocity profile $v(t)$.
\item Compute the derivatives ${\rm d}A(f)/{\rm d}f$, ${\rm d}\phi(f)/{\rm d}f$, and ${\rm d}^2\phi(f)/{\rm d}f^2$ with respect to $f$. 

\item Compute the distance profile  $d(t)\!=\!\int^{t}v(t')\,\textrm{d}t'$ and the acceleration profile $a(t)= \textrm{d} v(t) / \textrm{d}t$.

\item Compute the leading order corrections to the amplitude and phase:
\begin{align}
\delta\phi(f)&=-2\pi f \,d\left( \frac{{\rm d}\phi/{\rm d}f}{2\pi} \right)\,,
\\
\delta A(f)&=- \left[A(f)+f\frac{{\rm d}A}{{\rm d}f}\right] \, v\left( \frac{{\rm d}\phi/{\rm d}f}{2\pi} \right) 
\nonumber \\ &\quad\;- \frac{f A(f)}{4\pi} \frac{{\rm d}^2 \phi }{{\rm d}f^2}  \,a\left( \frac{{\rm d}\phi/{\rm d}f}{2\pi} \right);
\end{align}
\item The Doppler-shifted frequency-domain waveform model is then given by 
\begin{align} \label{eq:FINAL_FD_DOPPLER_RESULT}
\tilde{h}_{\rm D}(f) = [A(f)\!+\!\delta A(f)] \times \exp\!\Big\{i \big[\phi(f) \!+\! \delta\phi(f)\big]\Big\}.
\end{align}
\end{enumerate}

\subsection{Approximations}

In deriving the result in Eq.~\eqref{eq:FINAL}, the only two approximations that have been made are the SPA and a linear-order expansion in $\varepsilon$, which are deeply connected. %
Taking the $\varepsilon$ expansion to second order would require going beyond the SPA, which is a difficult but surmountable technical challenge (see e.g.~\cite{1999PhRvD..59l4016D}).
The leading-order approximation in $\varepsilon$ is expected to work as long as the projected peculiar velocity $v(t)$ is much smaller than the speed of light. 
We anticipate one could reduce the error from this approximation through a resummation technique, such as iterating over multiple stages of Doppler shifts of increasing velocity or using a Pad\'e fraction; this will not be pursued in this paper because the accuracy of the method at linear order is already probably sufficient for most future observations with third-generation detectors.

The SPA is expected to work as long as the phase varies much more rapidly than the amplitude, thus breaking down near the merger. In particular, the SPA allows us to write $t_{\rm s} \propto {\rm d}\phi/{\rm d}f$, but, near merger, the phase derivative fails as a proper ``clock''. There is a set of times near and after merger at which ${\rm d}\phi/{\rm d}f \leq 0$ and
the SPA time $t_{\rm s}$ does not advance forward (see e.g. Fig.~5 in~\cite{2016PhRvD..93d4007K}). In other terms, at late times the phase of the emitted GWs is not a good clock to parametrize the waveform signal. As we will show below, the effect of the breakdown of the SPA has, in practice, a very minor impact when comparing waveform models. We have tested various high-frequency extensions of the phase derivative to better model the SPA time and found negligible improvements over the simpler treatment presented here. This is mostly because very little signal-to-noise ratio is contained at those high frequencies.

\begin{figure*}
\centering
\includegraphics[width=0.72\textwidth]{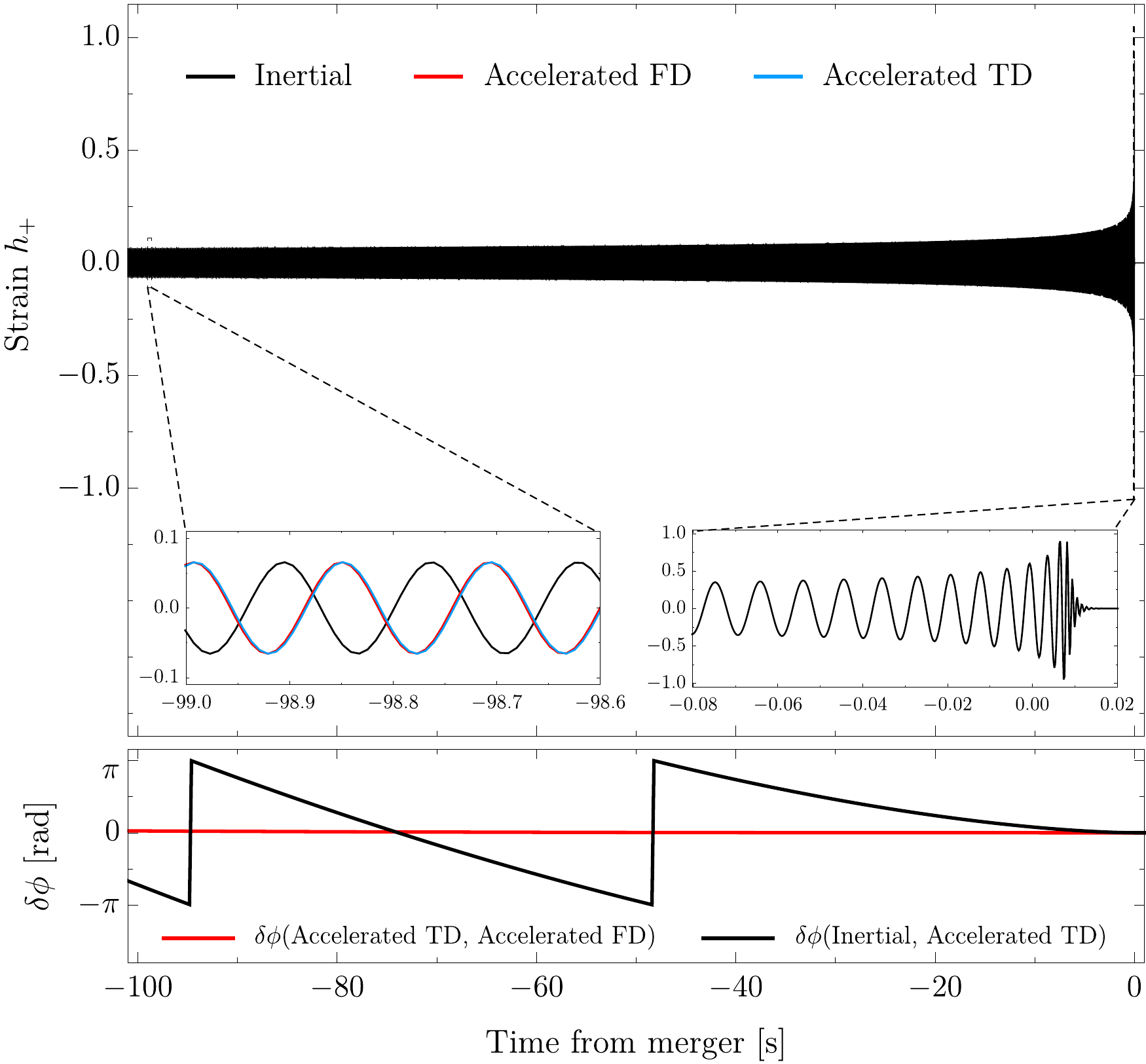}
\caption{Gravitational wave strains in the time-domain for an equal mass $m_1 = m_2 = 10 \, M_\odot$ non-spinning system as viewed from (i) an inertial frame at rest with respect to the binary (black curve), (ii) an accelerating frame with $a=10^{4} a_{0}$ computed exactly in the time domain (blue curve; TD), and (iii) the same accelerating frame waveform computed with the frequency-domain expressions developed in this paper (red curve; FD). The accelerated frame is chosen such that it coincides with the rest frame at the instant of merger. Accelerated and inertial waveforms, therefore, are in phase near merger and drift out of phase in the early inspiral (this can be seen most clearly in the inset plots). The bottom panel shows the (wrapped) phase difference between different pairs of waveforms. The black curve shows the phase difference between the accelerated TD waveform and the inertial frame waveform; observe that they dephase by just over one complete cycle in the 100 seconds before merger. The red curve shows the phase difference between the TD and FD methods of computing the accelerated waveform; the fast FD method dephases from the exact TD method by less than 0.1 radians over this time interval. 
}
\label{ConstantAccelerationFig}
\end{figure*}

\begin{figure}
\centering
\includegraphics[width=0.49\textwidth]{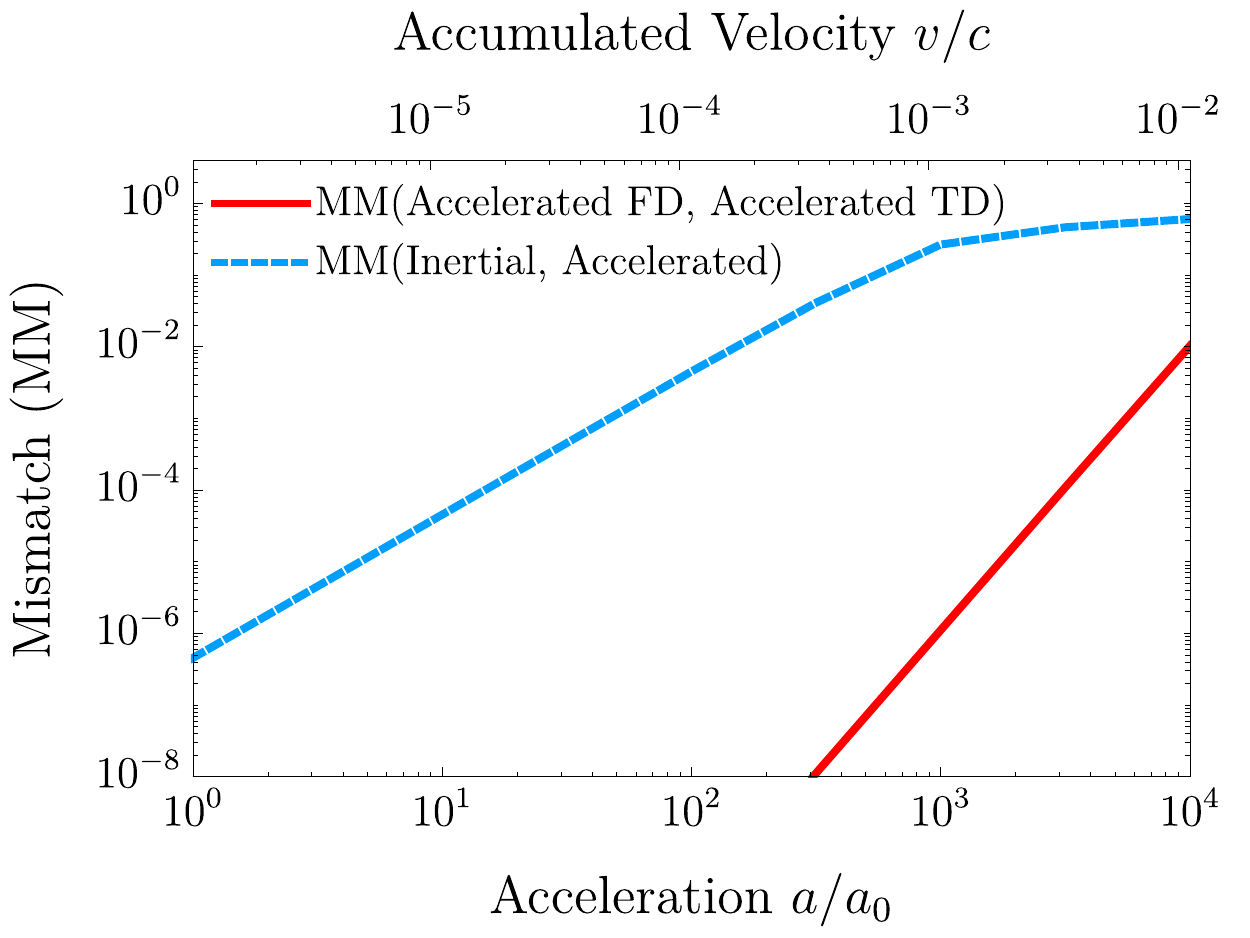}
\caption{Mismatch between the inertial and accelerated waveforms (blue dashed curve), and between the accelerated TD and accelerated FD waveforms (solid red curve) as a function of the magnitude of the acceleration or the accumulated velocity. Mismatches are computed using the ET-D noise PSD~\cite{2011CQGra..28i4013H} starting from $5\,\textrm{Hz}$, for which the signal duration is $T=240\,\textrm{s}$. During this time, the accelerated frame accumulates a total change in velocity of $v=aT$ relative to the inertial frame as reported on the upper $x$-axis. Even at large values of the acceleration, the fast frequency-domain approximation shows excellent agreement with the exact time-domain method: mismatches are less than $10^{-2}$ in all cases and improve rapidly as the acceleration decreases to more astrophysically realistic values.}
\label{AccelerationOverlapFigure}
\end{figure}

\begin{figure*}
\centering
\includegraphics[width=0.72\textwidth]{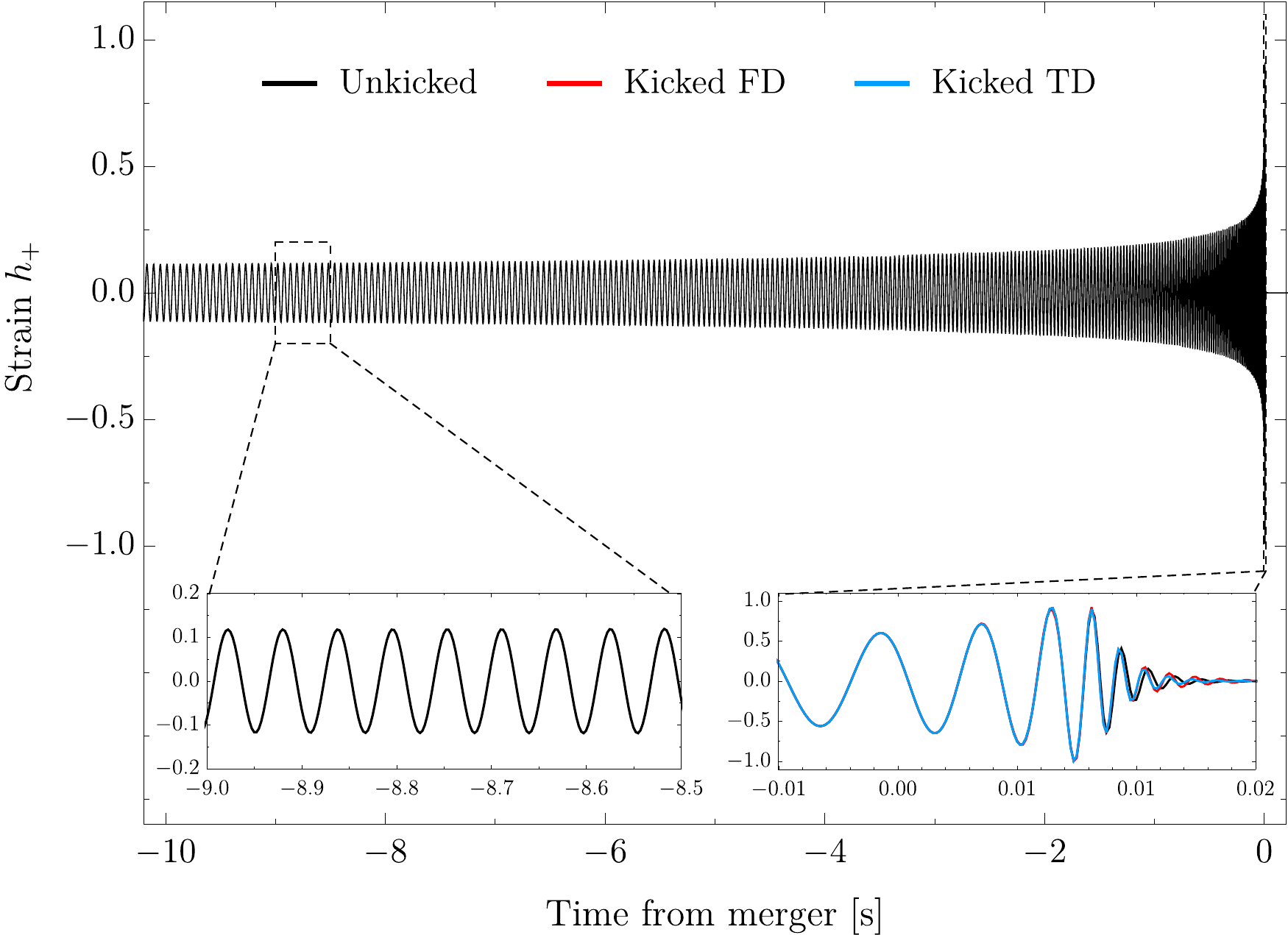}
\caption{Plotted in black in the main panel is the normal, ``unkicked'' waveform for an equal mass $m_1 = m_2 = 10 \, M_\odot$ non-spinning waveform. The two coloured curves show the same waveform but with an artificial recoil kick of $v_k=0.1c$ (unphysically large for testing purposes) applied at merger computed exactly in the time-domain (blue; TD) and the  frequency-domain approximation developed in this paper (red; FD). 
The kick is only imparted for a period about $\sigma=10 (m_1+m_2)$ near merger. The kicked and unkicked waveforms are therefore in phase in the early inspiral and drift out of phase during the merger and ringdown phase.
The two methods of calculating the kicked waveform signal (TD and FD) are in excellent agreement even for such a large value of the kick velocity. The difference between the two is barely visible in the ringdown signal in the right hand inset plot.}
\label{KickFig}
\end{figure*}

\section{Applications and tests}
\label{sec:applications}

In this section the frequency-domain method described in Sec.~\ref{recipe} is applied to two astrophysically motivated situations where Doppler shifted GW signals can be expected to occur. 
Section \ref{subsec:constacc} considers a merging stellar mass BH binary accelerated relative to a distant observer by the gravitational field of a nearby supermassive BH. Section \ref{subsec:kick} considers the acceleration a merging binary can impact on itself via a merger recoil, or ``kick''.

In both cases our primary focus will not be the astrophysics giving rise to the acceleration, but rather, it will be on testing the frequency-domain method described above in Sec.~\ref{recipe} by demonstrating that it correctly describes the Doppler shifted gravitational waves.
This will be done by comparing against a time domain method which explicitly includes the desired Doppler shifting.
For clarity, the procedure for performing the Doppler shift in the time domain is given here explicitly:
\begin{enumerate}
\item Evaluate $\tilde{h}(f)$ using Eq.~(\ref{eq:FDwaveform}) to obtain a numerical frequency domain waveform; ${\tilde{H} \! = \! {\big\{[j\Delta f,\tilde{h}(j\Delta f)\big]\,\big|\,j\!=\!0,1... n\big\}}}$.%
\item Window the numerical waveform below the lower starting frequency range that will be used to compute the match. 
\item Evaluate the numerical inverse FT using a standard Fast-Fourier-Transform  algorithm to obtain the time domain waveform $H \!=\! {\big\{ [j\Delta t,h(j\Delta t)]\,\big|\,j\!=\!0,1... n\big\}}$.
\item Create an interpolant of the numerical time domain waveform $\mathcal{H}(t)$.
\item The numerical Doppler-shift time-domain waveform is then obtained by evaluating this interpolant at the redshifted times from Eq.~(\ref{eq:redshifttime}), i.e. ${H_{\textrm{D}} \!=\! {\big\{ [j\Delta t,\mathcal{H}(t_{\textrm{D}})]\,\big |\, j\!=\!0,1,\ldots n\big\}}}$.
\item Perform a numerical forward FT to obtain the numerical frequency-domain waveform $\tilde{H}_{\textrm{D}}$ to be compared against our analytical result.
\end{enumerate}
We have checked that the discretization and multiple Fourier transforms needed to generate this frequency-domain waveform with the above time-domain Doppler shifts do not introduce artificial numerical artifacts due to aliasing, windowing or other undesirable features.  

The input frequency-domain waveform model used in the examples is PhenomD~\cite{2016PhRvD..93d4006H,2016PhRvD..93d4007K}. We stress that our approach is entirely independent on the base waveform model and can be applied to any frequency-domain approximant.

\subsection{Binaries in external gravitational potentials}
\label{subsec:constacc}

First, we consider the case where a stellar-mass BH binary resides close (a distance $R$) to a large third body, such as a supermassive BH with mass $M$. The third body accelerates the binary, relative to a distant observer on Earth, at a rate
\begin{align} 
a \approx \frac{GM}{R^{2}} = a_{0}\left(\frac{M}{10^{9}M_{\odot}}\right)\left(\frac{10^{-2}\,\mathrm{pc}}{R}\right)^{2} \,,
\end{align}
where $a_{0}=1.39\,\mathrm{m}/\mathrm{s}^{2}$. We assume that this acceleration is directed away from the observer on Earth and that, for convenience, the relative velocity between the observer and the binary is zero at merger (this needn't be the case but it makes the interpretation of our results easier). The merger time $t_0$ is taken to be equal to the coalescence time of the underlying PhenomD model ($t_c$ in the notation of~\cite{2016PhRvD..93d4006H,2016PhRvD..93d4007K}). Therefore, we have the following explicit expressions for the acceleration, velocity and displacement of the binary;
\begin{align}
a(t) &\equiv a = \mathrm{constant}, \label{eq:sys_1} 
\\ v(t) &= a\times(t-t_{0}), \label{eq:sys_2}
\\
 d(t) &= \frac{a}{2} \times (t-t_{0})^{2} \,. \label{eq:sys_3}
\end{align} 
Of course, the third-body acceleration will not remain exactly constant during the inspiral.
If $r_{12}$ is the orbital separation of the compact binary, then there will exist differences in the accelerations of the two objects (tidal accelerations) at the level ${\cal{O}}(r_{12}/R)$; this ratio is less than $10^{-9}$ when $r_{12} < 10^{3}\,\mathrm{km}$ and $R > 10^{-2}\,\mathrm{pc}$. 
Furthermore, as the binary orbits the supermassive BH the component of the acceleration along the line of sight will change by $\mathcal{O}(t_{\rm obs}/t_{\rm orbit})$, where $t_{\rm obs}$ is the duration of the GW signal and $t_{\rm orbit}\approx M (R/M)^{3/2}$). This ratio is less than $10^{-6}$ when $R > 10^{-2}\,\mathrm{pc}$, $M\approx 10^{9}\,M_{\odot}$, and for a typical LIGO/Virgo source with $t_{\rm obs} < 10^{2}\,\mathrm{s}$.

For a constant acceleration, the longer the signal lasts the greater the dephasing effect of the Doppler shift is. For a merging binary that is quasi-circular, with equal masses (here we pick $m_{1}=m_{2}= 10\,M_{\odot}$) and zero spins, the effect is then maximized when the detector's sensitivity curve can reach as low a frequency as possible. We thus imagine that this system is observed by the Einstein Telescope, with a lower starting GW frequency of $5\,\mathrm{Hz}$. %

The GW signal for one system with a large acceleration of $a=10^{4}a_{0}$ is shown in Fig.~\ref{ConstantAccelerationFig}. There are three curves in this figure corresponding to the waveform as viewed in its rest frame (inertial), the accelerated waveform as computed using the frequency-domain method described in Sec.~\ref{calculations} (Accelerated FD), and the time-domain method described earlier in Sec.~\ref{sec:applications} (Accelerated TD). Both accelerated waveform gradually dephase from the inertial waveform, as would be expected, with a total dephasing of just over a full cycle in the final 100 seconds of the inspiral. The accelerated FD waveform, however, remains closely in phase with the accelerated TD, with a dephasing of less than $0.1$ radians in the same amount of inspiral time. 

As described in Sec.~\ref{recipe}, we expect that the FD method will work best when the Doppler shifting velocity is small. Figure~\ref{AccelerationOverlapFigure} confirms this expectation by showing the waveform mismatch as a function of the magnitude of the acceleration (or, equivalently, as a function of the total change in velocity between the source and observer during the observation). The overlap between two waveforms $h_{1}$ and $h_{2}$ is here defined in the usual way
\begin{align}
{\rm{MM}} &= 1 - \max_{\phi,t}\frac{(h_{1}|h_{2})}{\sqrt{(h_{1}|h_{1})(h_{2}|h_{2})}}\,,
\end{align}
where the maximization is over an overall phase and time offset 
 and the signal inner product is defined by
\begin{align}
(h_{1}|h_{2}) &= 2 \int\; \frac{\tilde{h}_{1}^{*}(f) \tilde{h}_{2}(f) + \tilde{h}_{1}(f) \tilde{h}_{2}^{*}(f)}{S_{n}(f)} {\rm d} f\,,
\end{align}
where is
$S_{n}$ the spectral noise density curve of the detector (here taken to be that of ET-D with a starting low frequency of $5$ Hz~\cite{2011CQGra..28i4013H}). Observe that the mismatch between the accelerated TD and the accelerated FD model remains below 1\% for all $a_{0}$ considered, while the mismatch between the inertial and the accelerated TD model grows to to $\mathcal{O}(1)$. 

The parameter $t_{0}$ in Eqs.~(\ref{eq:sys_1}-\ref{eq:sys_3}) controls the time at which relative velocity between the observer and the binary vanishes.
Changing $t_{0}$ corresponds to the addition of a constant velocity offset between the source and observer which is degenerate with the source's total mass.
If $t_{0}$ is chosen to correspond to some point in the early inspiral (instead of near merger as was done above), then the signal has longer to drift out of phase and larger mismatch values are obtained. However, most of this larger mismatch will be absorbed into the measurement of a redshifted value of the total mass, leaving behind the mismatches reported in Fig.~\ref{AccelerationOverlapFigure}.

\subsection{Black-hole merger recoils}\label{subsec:kick}

We now consider Doppler shifts that resemble binary BH recoil merger kicks~\cite{2016PhRvL.117a1101G,2009JPhCS.154a2043F} (other observables of BH kicks have been proposed, see e.g.~\cite{2018arXiv180611160C}). For simplicity, we assume the acceleration profile that resembles a kick is a Gaussian of constant width, centered at the time of merger, and we just vary the final velocity, $v_k$, of the remnant BH:
\begin{align}
a(t) &= \frac{v_k}{\sqrt{2 \pi} \sigma} \exp\left[{-\frac{1}{2} \left( \frac{t-t_0}{\sigma}\right)^2 } \right]
\\
v(t) &= \frac{v_k}{2} \left[1+ \erf \left(\frac{t-t_0}{\sqrt{2}\sigma}\right) \right], \\ 
d(t) &= \sigma^{2} a(t) + (t-t_0) v(t)\,.
\end{align}
where $\sigma$ and $t_0$ are two parameters describing the timespan over which the kick is imparted and its center. We set $\sigma=10M$, as was found to be a good approximation in careful comparisons against numerical relativity simulations~\cite{2018PhRvD..97j4049G}. The kick center $t_0$ is set using the SPA time corresponding to the last amplitude transition frequency of PhenomD (i.e. $f_{\rm peak}$ in the notation of~\cite{2016PhRvD..93d4006H,2016PhRvD..93d4007K})
The merging binary is taken to be a quasi-circular, equal mass, $m_{1}=m_{2}=10\,M_{\odot}$, non-spinning system. In this case the effect of the kick is confined to be close to merger, so it is not necessary to have a very long signal. Therefore, we imagine that this system is observed by Advanced LIGO at design sensitivity (PSD from~\cite{LIGOpsd} with a lower starting frequency of $30\,\mathrm{Hz}$).

\begin{figure}
\centering
\includegraphics[width=0.49\textwidth]{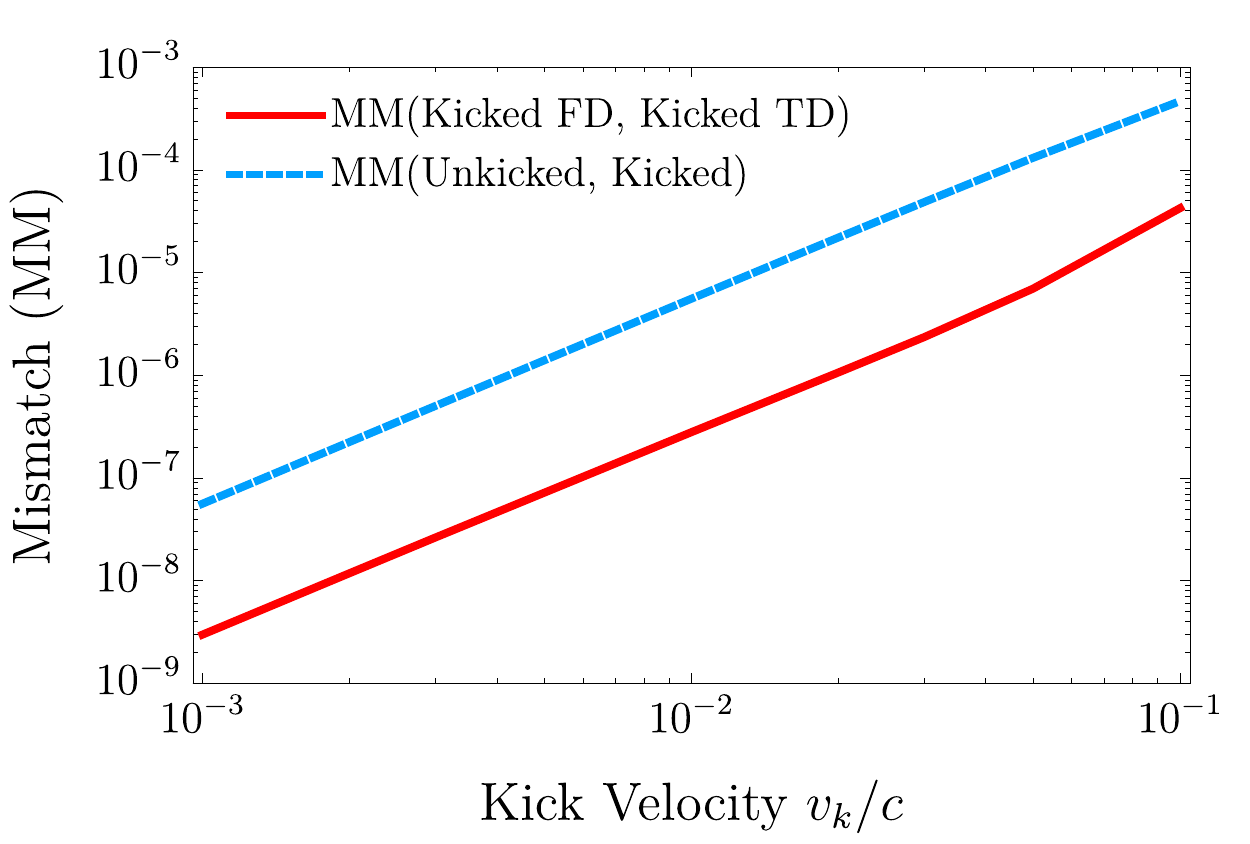}
\caption{Mismatch between the kicked and inertial waveforms (blue dashed curve), and between the kicked TD and kicked FD waveforms (solid red curve) as a function of the magnitude of the recoil kick velocity. Mismatches were computed using the LIGO noise PSD~\cite{LIGOpsd} starting from $30\,\textrm{Hz}$. Even at large values of the recoil velocity, our frequency-domain recipe shows excellent agreement with the exact time-domain method. Mismatches improve exponentially as the velocity decreases to more astrophysically realistic values.}
\label{KickOverlapFigure}
\end{figure}

In reality, kick velocities are at most of $\mathcal{O}(10^{-2} c)$. %
The GW signal for one system with an unphysically large kick velocity $v_k=0.1c$ away from the observer is shown in Fig.~\ref{ConstantAccelerationFig}. Again, there three curves in this figure, corresponding to the original waveform without any artificial kick (unkicked), the kicked waveform as computed using the frequency-domain method described in Sec.~\ref{calculations} (Kicked FD), and the time-domain method described in Sec.~\ref{sec:applications} (Kicked TD). As expected, the kicked waveforms dephase from the inertial waveform during the merger. The accelerated FD and TD waveforms remain close in phase even during the late ringdown. As described above, we expect that the FD method will work best when the Doppler shifting kick velocity is small. Figure~\ref{KickOverlapFigure} confirms this expectation by showing the waveform mismatch as a function of the magnitude of the kick velocity. Our FD procedure well reproduces the TD result, with mismatches which are over an order of magnitude smaller than those due to the kick.

\section{Conclusions}

This paper presents advances in gravitational waveform building to model effects that impact the signal at the level of a Doppler shift. In particular, we present a frequency-domain method that takes a gravitational waveform model constructed in an inertial frame and lifts it to an accelerated frame through the addition of a frequency-dependent amplitude and phase correction. This method is fast and straightforward to implement, requiring only knowledge of the time-dependent velocity profile and the frequency-domain gravitational wave amplitude and phase. Moreover, the method is faithful, resulting in matches well above 99\% for the cases investigated here.

The method developed here, therefore, is ready to be implemented in data analysis investigations. One possible future use is to study how well the presence of a third body, through its acceleration of the center of mass of the GW-emitting binary, can be determined with future observations with third-generation detectors and space-borne detectors. Similar studies could be carried out on the McVittie effect due to the accelerated expansion of the universe~\cite{McVittie:1959afu}, as well on modified gravity effects that lead to an acceleration of the center of mass~\cite{2010PhRvD..81f4018Y}. One could for example investigate the signal-to-noise ratio that would be required to extract these effects from the GWs emitted by coalescing compact binaries as a function of future detectors, which in turn, could provide guidance toward instrument design.  

Our frequency-domain method could also be improved to obtain a more accurate representation of the acceleration of the center of mass of GW emitting binaries. The method does rely on an expansion of the center-of-mass velocity relative to the speed of light, which we have carried out here to leading order. An extension of this method to higher order would be straightforward, although it would also require going beyond the SPA, including the next-to-leading order terms in the method of steepest descent~\cite{1999PhRvD..59l4016D}. Moreover, one could also in principle carry out a resummation of the expanded amplitude and phase corrections, for example through a Pad\'e approximant. Our match calculation, however, suggests that any such improvements may only be necessary for the highest signal-to-noise ratio events of third-generation detectors. 

\label{end}
\vspace{-15pt}
\acknowledgements 
\vspace{-5pt}
We Riccardo Barbieri, Ulrich Sperhake, Ron Tso and Kaze Wong for fruitful discussions. 
K.C. acknowledges support from the LIGO
SURF program at Caltech  through NSF Grant No. PHY-1460838. 
D.G. is supported by NASA through Einstein Postdoctoral Fellowship
Grant No. PF6-170152 by the Chandra X-ray Center,
operated by the Smithsonian Astrophysical Observatory
for NASA under Contract NAS8-03060. 
C.J.M. acknowledges financial support from European Union's H2020 ERC Consolidator Grant ``Matter and strong-field gravity: New frontiers in Einstein's theory'' grant agreement No.\ MaGRaTh--646597 and European Union's  H2020 research and innovation programme under the Marie Sklodowska-Curie grant agreement No.\ 690904.
N.Y acknowledges support from NSF CAREER Grant No. PHY-1250636 and NASA Grants NNX16AB98G and 80NSSC17M0041.

\vspace{-5pt}
\bibliographystyle{apsrev4-1}
\bibliography{fdkick}

\end{document}

